\documentclass[fleqn,usenatbib]{mnras}

\usepackage{newtxtext,newtxmath}
\usepackage[dvipsnames]{xcolor}

\usepackage[T1]{fontenc}
\usepackage{float}
\DeclareRobustCommand{\VAN}[3]{#2}
\let\VANthebibliography\thebibliography
\def\thebibliography{\DeclareRobustCommand{\VAN}[3]{##3}\VANthebibliography}

\usepackage{soul}
\usepackage{ulem}
\usepackage{tabu}

\usepackage{graphicx}	
\usepackage{amsmath}	
\usepackage{xcolor}

\usepackage{hyperref} 
\title[GW events without GRBs]{On non-detection of Gamma-Ray Bursts in three compact binary merger events detected by LIGO}

\author[Mazwi, Razzaque \& Nyadzani]{
Luyanda Mazwi,$^{1}$\thanks{E-mail: luyandamazwi10@gmail.com}
Soebur Razzaque,$^{1,2,3}$\thanks{E-mail: srazzaque@uj.ac.za}
Lutendo Nyadzani,$^{1}$\thanks{E-mail: lnyadzani@uj.ac.za}
\\
$^{1}$Centre for Astro-Particle Physics (CAPP) and Department of Physics, University of Johannesburg, PO Box 525, Auckland Park 2006, South Africa\\
$^{2}$Department of Physics, The George Washington University, Washington, DC 20052, USA \\
$^{3}$National Institute for Theoretical and Computational Sciences (NITheCS), South Africa 
}

\date{Accepted XXX. Received YYY; in original form ZZZ}

\pubyear{2015}

\begin{document}
\label{firstpage}
\pagerange{\pageref{firstpage}--\pageref{lastpage}}
\maketitle

\begin{abstract}
The joint detection of the gravitational wave (GW) event GW170817 and the short-duration gamma-ray burst (SGRB) event GRB~170817A, marked the beginning of GW multi-messenger astronomy and confirmed that binary neutron star mergers are progenitors of at least some SGRBs. An estimated joint detection rate of 0.3 -- 1.7 per year between the LIGO-Hanford, LIGO-Livingston and Virgo GW network at design sensitivity, and the {\it Fermi} Gamma-ray Burst Monitor was predicted. However, to date, the GW170817/GRB~170817A joint detection has been the only event of its kind so far. Taking into account that SGRBs are narrowly beamed and are emitted perpendicular to the orbital plane of the binary system, we propose that previous mergers involving neutron stars, were orientated such that observation of the emitted SGRB along this narrow jet was not possible. To support this hypothesis we have estimated the inclination of the binary systems for previously detected Binary Neutron Star (BNS) and Black Hole Neutron Star (BHNS) mergers through GW analysis. This analysis was performed using \textsc{BILBY}, a Python based Bayesian inference library, to estimate the inclination of the BNS events GW170817 and GW190425, and the BHNS events GW190917\_114630 and GW200115\_042309. The results obtained in this study indicate that these binaries may have had inclinations greater than $33^{\circ}$ with respect to the line of sight from Earth, an upper limit on the viewing angle set from observations of GRB~170817A. This then suggests that the observation of the emitted SGRB from these past mergers might not have been possible. 
\end{abstract}

\begin{keywords}
gravitational waves - (stars:) gamma-ray burst: general - methods: data analysis
\end{keywords}



\section{Introduction}
Gravitational waves (GWs) are travelling disturbances in spacetime that are caused by the acceleration of massive bodies. These ripples travel away from the source at the speed of light, containing information about their source \citep[see, for review,][]{sathyaprakash2009physics}. Despite being one of the key predictions of General Relativity, the first indirect observational evidence of their existence was from observations of the orbital decay of the Hulse and Taylor binary pulsar PSR 1913+16 due to the emission of GWs \citep{1991ApJ...366..501D,1982ApJ...253..908T}. It was only on 14 September 2015 that the first direct observation of GWs by the Laser Interferometer Gravitational wave Observatory (LIGO) was achieved during its first observing run (O1). GW150914 was detected from the merger of two black holes in a binary, marking the beginning of the GW Astronomy era \citep{abbott2016observation}. By studying these waves we are then able to study the motion and macroscopic behaviour of these systems.

Gamma-ray bursts (GRBs) are bursts of non-thermal gamma radiation (typically, 1 keV -- 1 MeV photons) which are extra-galactic in origin. First discovered in the late 1960s, these phenomena have long puzzled astronomers. GRB sources are extremely luminous, emitting photons with a total energy of $\sim 10^{51} - 10^{53} ~\textrm{ergs}$ in a few seconds, making them one of the most luminous electromagnetic (EM) phenomena in the universe \citep[see, for reviews,][]{PhysicsOfGRBS, Meszaros2006RPPh...69.2259M, Gehrels2013FrPhy...8..661G, Kumar2015PhR...561....1K}. They can be classified as either long or short GRBs with these two classes arising from distinct physical phenomena \cite[see, e.g.,][]{Kouveliotou1993ApJ...413L.101K, nakar2007short}. In particular, the sources of SGRBs remained unknown for a long time, despite the most likely candidates being binary neutron star (BNS) or Black Hole Neutron Star (BHNS) mergers. On 17 August 2017 at 12:41:04 UTC, the first GW event produced from a BNS merger, GW170817, was detected by the LIGO-Hanford and LIGO-Livingston detectors. Approximately 1.7~s later, the SGRB event GRB~170817A was detected independently by the {\it Fermi} Gamma-ray Burst Monitor (GBM), and the Anti-Coincidence Shield for the Spectrometer for the International Gamma-Ray Astrophysics Laboratory (INTEGRAL)\citep{abbott2017gravitational}. The probability of the almost simultaneous spatial and temporal observation of these two events being due to chance coincidence is $5\times 10^{-8}$. We can then confirm that BNS mergers are the progenitors of at least some SGRBs with this event marking the beginning of GW multi-messenger astronomy \citep{abbott2017gravitational}.

From the observations of the GW/SGRB events of 17 August 2017, a joint detection rate of 0.1 -- 1.4 per year between the LIGO and the {\it Fermi}-GBM during the third observing run (O3) of the LIGO/Virgo Collaboration (LVC) was predicted, and a joint detection rate of 0.3 -- 1.7 per year was predicted at the design sensitivity \citep{abbott2017gravitational}. However, O3 has been completed with no other joint detection. LIGO detected one more BNS merger and possibly four BHNS merger events without coincident SGRB detection during O3. This study aims to uncover a possible reason, based on orbital inclination angle, for the lack of further GW and SGRB joint detection.

If we make the naive assumption that all merger events involving a neutron star produce an SGRB, a few explanations for the lack of joint detection arise. When considering our ability to observe GRBs, the limitations of gamma-ray instruments is a probable cause for the lack of EM counterpart observations for the past GW events. Subluminous GRB events like GRB 170817A can only be detected out to a distance of about 80 Mpc before falling below the detection threshold of instruments like {\it Fermi}-GBM \citep{abbott2017gravitational}. Therefore, other GW events similar to GW170817 may not have an observable counterpart if they occur at farther distances. Secondly, gamma-ray instruments only provide partial sky coverage. Depending on the sky location of the GW event, complete sky coverage in that direction may not be possible. For example {\it Fermi}-GBM and {\it Swift}-BAT only had a combined sky coverage of $57.81 \%$ during GW190425 while the coverage was much higher for GW190917\_114630 ($95\%$) and GW200115\_042309 ($96\%$) \citep{Fletcher_2024}. 
From the perspective of GW astronomy, two possible reasons for the lack of joint detection arise. The first being that the merger event was outside the horizon distance (which is the furthest a source can be detected above a signal-to-noise ratio threshold of 8 \citep{allen2012findchirp}) of the current ground based GW interferometers. In this instance, a GRB may be detected but the GW counterpart may be undetectable. The second possibility has to do with the orientation of the binary system that produces the GW and EM emission. GRBs are narrowly beamed, with a viewing angle of $1/\Gamma$, where $\Gamma$ is the Lorentz factor of the relativistic GRB jet. However, matter moves relativistically in the processes that produce GRBs meaning $\Gamma$ is very large. This means the GRB has a very narrow viewing angle. Most GRBs have typical jet opening angles of $2^{\circ} \lesssim \theta \lesssim 20^{\circ}$ \citep[see, e.g,][]{goldstein16}.  According to the current models for the production of SGRBs, after the BNS merger, an accretion disk is formed around the merged object, either a black hole or a neutron star, and the GRB jet is launched perpendicular to the orbital plane of the binary system along its rotational axis \citep[see, e.g.,][]{Baiotti2017RPPh...80i6901B}. We assume that this also holds true in the case of a BHNS merger. Therefore, observation of an SGRB from a BNS or BHNS merger event will only be possible if the binary system is orientated almost face-on with Earth. In this study we focus on this explanation for the lack of joint detection and show that only GW170817 had an orientation such that observation of an SGRB was possible.

We organise this paper in the following order. In Sec.~2 we present our analysis of GW data from LIGO for four merger events involving at least one NS in the binary system. We present our results in Sec.~3. We discuss our findings in Sec.~4 and conclude in Sec.~5.

\section{Analysis of LIGO data from Run O3}
Gravitational waves can be accurately modelled through analytical or numerical solutions to the Einstein equations \citep{ajith2011addressing}. These waveform models are parameterised by several parameters such as the component masses and spins, which provide information about the progenitor systems. By analysing the GW data gathered by the LIGO interferometers, it is possible to infer the values of these parameters. We perform this inference through the framework of Bayesian analysis as the detected signal is not precisely known, the detector noise is not perfectly modelled and because GWs are so weak, resulting in large uncertainties in the measurements of these values \citep{abbott2020guide}. To perform this analysis, we make use of \textsc{BILBY},\footnote{ \url{https://pypi.org/project/bilby/}} a python based Bayesian inference library that was developed for GW analysis by \citet{ashton2019bilby}. Before performing the analysis, a basic treatment to the data was applied. This was done for each GW transient event that was analysed in this study. The Power Spectral Density (PSD) was modelled for 128~s around the GW events to provide a model of the detector noise. For short periods (a few seconds to a few minutes), the detector noise is roughly stationary \citep{abbott2020guide}. This PSD was estimated using the median average of overlapping periodograms. In the instance of GW170817, a short noise transient (glitch) of unknown origin is present in the LIGO-Livingston detector data. In this study, we made use of the cleaned data around this event, where the glitch was modelled and subtracted from the time series data.\footnote{\url{https://dcc.ligo.org/LIGO-T1700406/public}} All other data used in this study is publicly available online from LIGO.\footnote{\url{https://gwosc.org/eventapi/html/}}

\subsection{Selection of events}
\label{sec:events} 
In this study, we focus on GW events with $p_{\mathrm{astro}} > 0.5$ so as to have a higher probability of the signal being of astrophysical origin. We studied two BNS events (GW170817 and GW190425) and two BHNS events (GW190917\_114630 and GW200115\_042309). We analyse GW170817 in order to compare the results we obtained by \textsc{BILBY} with that by the LIGO Collaboration. Due to the detection of an EM counterpart to GW170817 (i.e., GRB~170817A), this event was well localised and the EM constraint on the luminosity distance to the source enabled a fairly accurate measurement of the inclination of the binary. When performing the analysis of GW170817, we purposefully specified as little information as possible so as to mimic other events that are not so well localised.

GW190425, is the only other detected BNS merger event. When considering which BHNS merger events from O3 to include in our study, only GW190917\_114630 and GW200115\_042309 (from now on abbreviated as GW190917 and GW200115) were selected. There are two other BHNS events, namely GW191219\_163120 and GW200210\_114636, however they were excluded from this study. This is because for a system with mass ratio $q$, $q \ll 0.1$, the neutron star falls into the black hole without accreting, a necessary condition for the production of a GRB \citep{nakar2007short}. Secondly, simulations for a neutron star with a polytropic equation of state with a polytropic index of 1, no accretion disk was formed for BHNS systems with a mass ratio of $0.1$ \citep{faber2006general}. GW191219\_163120 and GW200210\_114636 have mass ratios $q \approx 0.0363$ and $q \approx 0.1059$, respectively. We can then argue that no GRB would have been formed during these merger events. Lastly, 
GW200105\_162426 
is excluded from this study despite it meeting the mass ratio condition meriting its inclusion, because $p_{\mathrm{astro}} = 0.36$ for this event and we cannot confidently conclude that it is of astrophysical origin. 

\subsection{Selection of waveform models}
For the analysis performed in this study, we used a set of frequency domain GW wave forms. For a binary system that is well separated and moving slowly $v/c \ll 1$ these wave forms can be computed using the post-Newtonian (PN) approximation to general relativity. However, once these bodies start moving relativistically, these approximations break down and accurate solutions to the Einstein equations are required to produce an accurate image of the merger process \citep{ajith2011addressing}. These solutions, however, can only be generated through numerical simulations. The waveform models used on the BNS merger signals were the {\it IMRPhenomPv2 NRTidal}, {\it IMRPhenomD NRTidal} and {\it TaylorF2} models and the models used on the BHNS merger events were {\it IMRPhenomPv2} and {\it IMRPhenomXPHM}.

The {\it TaylorF2} waveform model is an analytical PN model for GWs from non-spinning binaries in the quasi-circular inspiral phase in the frequency domain \citep{huerta2014accurate}. This model includes corrections up to 3.5 PN and is computed in the stationary phase approximation (SPA). The remaining 3 wave forms are all Inspiral Merger Ringdown (IMR) based on phenomenological (Phenom) treatments of the IMR. {\it IMRPhemomD} is a model based on aligned spin point particle models tuned to Numerical Relativity (NR) hybrids and Effective One Body (EOB) wave forms \citep{abbott2019properties}. {\it IMRPhenomP} is a phenomenological waveform model that includes spin precession \citep{abbott2019properties}. {\it IMRPhenomXPHM} models GWs from a quasi circular precessing binary black hole \citep{pratten2021computationally}.

\subsection{Choice of parameters and priors }
\label{sec:priors}
The observed GW signal is governed by extrinsic and intrinsic parameters. The intrinsic parameters are parameters of the binary itself, i.e., the component masses, mass ratio, chirp mass and spins of the bodies. The extrinsic parameters are the set of parameters governing the location and orientation of the binary with respect to the observer, i.e., the luminosity distance, sky location and inclination of the binary with respect to the line of sight. This information (represented as a set $\theta$) can be expressed as a probability density function known as the posterior probability function $p(\theta|d(t))$, given some data $d(t)$. Using Bayes' theorem we can determine this function as the product of the likelihood $\mathcal{L}(d(t)|\theta)$, the probability of observing the waveform given the parameters described, and the prior probability density function that contains our prior knowledge of the signal. These posteriors are computed using \textsc{BILBY}'s native Markov Chain Monte-Carlo (MCMC) sampler in order to create credible intervals for these parameters by marginalising over all parameters except a few. In order to begin this analysis, we need to choose intervals for the parameters we're interested in. In this study, all parameters governing the source location of the GWs were kept as general as possible. Only the priors governing the component masses and chirp mass were specified.

We follow a similar approach to choosing priors for our analysis of the BNS events as in \citet{abbott2019properties}. For the component masses $m_1$ and $m_2$, we adopt the convention that $m_1 \geq m_2$. We assume a uniform distribution in the detector frame, i.e., $m^{det} = m(1 +z)$ where $z$ is the redshift and the component masses are such that $0.5~\mathrm{M_{\odot}} \leq m^{det}_1 \leq 8~\mathrm{M_{\odot}}$ and $0.4 ~\mathrm{M_{\odot}} \leq m^{det}_2 \leq 7.7 ~\mathrm{M_{\odot}}$. We assume a uniform distribution in detector chirp mass ($\mathcal{M}^{det}$) such that $1.184 ~\mathrm{M_{\odot}} \leq \mathcal{M}^{det} \leq 2.168 ~\mathrm{M_{\odot}}$. With regard to the BHNS events we assumed uniform distributions in the component masses such that for GW200115, $2.1 ~\mathrm{M_{\odot}} \leq m^{det}_{1} \leq 7 ~\mathrm{M_{\odot}}$ and $1.1 ~\mathrm{M_{\odot}} \leq m^{det}_{2} \leq 2.5 ~\mathrm{M_{\odot}}$. For GW190917, the component mass prior was chosen as $2.1 ~\mathrm{M_{\odot}} \leq m^{det}_{1} \leq 10 ~\mathrm{M_{\odot}}$ and $1.1 ~\mathrm{M_{\odot}} \leq m^{det}_{2} \leq 2.5 ~\mathrm{M_{\odot}}$. For both GW190917 and GW200115 a uniform prior on the chirp mass was chosen where $2 ~\mathrm{M_{\odot}} \leq \mathcal{M}^{det} \leq 10 ~\mathrm{M_{\odot}}$. This choice in priors was informed by the reported component and chirp masses of these events \citep{abbott2017gravitational, abbott2020gw190425, abbott2021observation, web:gwosc}. 

Following \citet{abbott2019properties} we set two priors for the dimensionless spin parameter $\chi$. As a result we have two sets of priors for a high spin case where $\chi$ has a uniform distribution such that $0 \leq \chi \leq 0.89$. This allows for the possibility of exotic binary systems. For the low-spin prior, $\chi$ is bounded by $0$ and $0.05$. When analysing the 2 BHNS merger events GW190917 and GW200115, we only considered the high-spin prior to account for the spin of the black hole. For all GW transients studied here, we assume a cosine distribution in the declination and a uniform distribution in right ascension from 0 to $2\pi$ with a periodic boundary condition. For GW170817 and GW190425, we assume uniform priors for their respective luminosity distances $D_L$ around the reported values. The prior we set for GW170817 is $25 ~\mathrm{Mpc} \leq D_L \leq 47 ~\mathrm{Mpc}$, and for GW190425 we set $104 ~\mathrm{Mpc} \leq D_L \leq 188 ~\mathrm{Mpc}$. For the 2 BHNS mergers in this study, these events are reportedly much further away than the two BNS events \citep{abbott2021observation, web:gwosc}. We assume a power-law distribution in the luminosity distance prior for these two events with a power-law index $\alpha = 2$. We set the prior on the luminosity distance for GW200115 such that $202 ~\mathrm{Mpc} \leq D_L \leq 352 ~\mathrm{Mpc}$, and for GW190917 such that $410 ~\mathrm{Mpc} \leq D_{L} \leq 1060 ~\mathrm{Mpc}$.

Lastly, we set priors on the inclination, largely varying from event to event. For GW170817, a uniform prior in inclination $\iota$ was chosen where $130^{\circ} \leq \iota \leq 160^{\circ}$. A uniform prior in inclination where $0^{\circ} \leq \iota \leq 90^{\circ}$ was chosen for GW190425. This prior was chosen based on an assumption that GWs are symmetric about the orbital plane of the binary. For the 2 NSHB events, these events had a weaker Signal-to-Noise-ratio (SNR) than the other events studied here (GW200115 had an SNR of 11.5 and GW190917 an SNR of 8.3 \citep{web:gwosc}). As a result, two priors on the inclination were chosen for both of them. A uniform prior where $0^{\circ} \leq \iota \leq 90^{\circ}$, and a prior with a sinusoidal distribution in inclination running from $0$ to $2\pi$.

\section{Results}
We present the results with the inferred values for the inclination, chirp mass and mass ratio for the GW events in this study. We present the chirp mass and mass ratio as a way to estimate how well the analysis was performed. Since these parameters are intrinsic to the binary and are reported by LIGO, any major deviations in these values (in particular the chirp mass) to those in literature would indicate that something went wrong in our analysis.

\subsection{GW170817}
The titular GW170817 was a three detector event with triggers in both the LIGO-Hanford, LIGO-Livingston and Virgo detectors, all in observing mode \citep{abbott2017gw170817}. This GW originated from a BNS merger with component masses $m_{1} = 1.46(+0.12,-0.30) \mathrm{M_{\odot}}$ and $m_{2} = 1.27(+0.09,-0.09) \mathrm{M_{\odot}}$ . With a rather high SNR of 32.4, this event was the loudest GW signal for that period of the LVC network's operation. The results of the analysis performed using three GW waveform templates that are given in Tables \ref{tab:GW170817_LS} and \ref{tab:GW170817_HS} for the low-spin and high-spin priors, respectively. Errors quoted in the form $x^{+y}_{-z}$ represents the median value ($x$), $95\%$ upper limit ($y$) and $5\%$ lower limit ($z$).

\begin{table}
    \centering
    \begin{tabular}{cccc}
    \hline
        Waveform & Inclination & Chirp mass & Mass ratio  \\
         & ($^\circ$) & ($\mathrm{M_{\odot}}$) &  \\
    \hline
        {\it TaylorF2} & $142.88^{+0.9}_{-0.8}$ & $1.19^{+0.0}_{-0.0}$ & $0.42^{+0.17}_{-0.03}$\\
        {\it IMRPhenomP} & $155.28^{+15.99}_{-18.57}$ & $1.20^{+0.0}_{-0.0}$ & $0.83^{+0.11}_{-0.11}$\\
        {\it IMRPhenomD} & $155.21^{+15.98}_{-18.56}$ & $1.20^{+0.0}_{-0.0}$ & $0.83^{+0.11}_{-0.11}$\\
    \hline
    \end{tabular}
   \caption{Results for GW170817 from the low-spin prior.}
   \label{tab:GW170817_LS}
\end{table}

\begin{table}
    \centering
    \begin{tabular}{cccc}
    \hline
        Waveform & Inclination & Chirp mass & Mass ratio  \\
         & ($^\circ$) & ($\mathrm{M_{\odot}}$) &  \\
    \hline
        {\it TaylorF2} & $152.41^{+18.65}_{-15.82}$ & $1.19^{+0.0}_{-0.0}$& $0.61^{+0.26}_{-0.23}$\\
        {\it IMRPhenomP} & $155.28^{+18.65}_{-16.01}$ & $1.19^{+0.0}_{-0.0}$& $0.65^{+0.24}_{-0.19}$\\
        {\it IMRPhenomD} & $155.57^{+15.62}_{-18.82}$ & $1.21^{+0.0}_{-0.0}$ & $0.69^{+0.19}_{-0.21}$\\
    \hline
    \end{tabular}
    \caption{Results for GW170817 using a high-spin prior.}
    \label{tab:GW170817_HS}
\end{table}

Interestingly enough, we find that the results obtained in our analysis is rather close to those obtained in \citet{abbott2019properties} using the EM constraint on the luminosity distance. The inclinations obtained there using the EM constraint are $151^{\circ} (+15^\circ, -11^\circ)$ and $153^{\circ} (+15^\circ, -11^\circ)$ for the low- and high-spin prior, respectively. The values obtained for the chirp mass for both the high- and low-spin priors in our analysis are comparable to the mass $1.19 ~\mathrm{M_{\odot}}$ found in \citet{abbott2019properties}. From these results, we conclude that \textsc{BILBY} is adequate for our analysis and that despite the lack of concrete information on the sky location, luminosity distance of the source, one could still obtain fair estimates of the inclination of these binaries. Of course there is the caveat that GW170817 had an extremely high SNR compared to the other events. 

\subsection{GW190425}
This is the second BNS detection. Being only a single detector event with an SNR of 12.9, with a total system mass of $3.4 ~\mathrm{M_{\odot}}$ and component masses $m_1 = 2.1(+0.5,-0.5) ~\mathrm{M_{\odot}}$ and $m_2 = 1.3(+0.3,-0.2) ~\mathrm{M_{\odot}}$ \citep{abbott2020gw190425}. This is the largest measured mass of a BNS system, implying the possibility that one of the component masses could be a black hole. The results for the inclination are given in Tables \ref{tab:GW190425_LS} and \ref{tab:GW190425_HS} for the low- and high-spin prior, respectively. For the analysis performed, data from the LIGO-Livingston and Virgo detectors were used as the LIGO-Hanford detector was offline for two hours around the event \citep{abbott2020gw190425}. No EM counterpart for this event was detected.

\begin{table}
    \centering
    \begin{tabular}{cccc}
    \hline
        Waveform & Inclination & Chirp mass & Mass ratio  \\
         & ($^\circ$) & ($\mathrm{M_{\odot}}$) &  \\
    \hline
        {\it TaylorF2} & $44.64^{+29.94}_{-28.45}$ & $1.47^{+0.02}_{-0.00}$ & $0.43^{+0.41}_{-0.09}$\\
        {\it IMRPhenomP} & $46.54^{+28.96}_{-30.54}$ & $1.47^{+0.02}_{-0.00}$ & $0.43^{+0.36}_{-0.07}$\\
        {\it IMRPhenomD} & $46.09^{+30.65}_{-32.38}$ & $1.47^{+0.02}_{-0.00}$& $0.43^{+0.36}_{-0.07}$\\
    \hline
    \end{tabular}
    \caption{Results for GW190425 using a low-spin prior.}
    \label{tab:GW190425_LS}
\end{table}

\begin{table}
    \centering
    \begin{tabular}{cccc}
    \hline
        Waveform & Inclination & Chirp mass & Mass ratio  \\
         & ($^\circ$) & ($\mathrm{M_{\odot}}$) &  \\
    \hline
        {\it TaylforF2} & $98.03^{+58.00}_{-63.39}$ & $1.49^{+0.01}_{-0.02}$ & $0.62^{+0.25}_{-0.33}$\\
        {\it IMRPhenomP} & $89.04^{+63.11}_{-60.36}$& $1.47^{+0.03}_{-0.00}$ & $0.35^{+0.43}_{-0.12}$\\
        {\it IMRPhenomD} & $87.95^{+62.85}_{-62.83}$ & $1.99^{+0.00}_{-0.00}$ & $0.13^{+0.00}_{-0.00}$\\
    \hline
    \end{tabular}
    \caption{Results for GW190425 using a high-spin prior.}
    \label{tab:GW190425_HS}
\end{table}
The chirp masses obtained in this analysis for the low-spin prior is comparable to that obtained by LIGO. This value in literature is given to be $1.44\pm 0.02 ~\mathrm{M_{\odot}}$ \citep{abbott2020gw190425}. The values obtained for the high-spin prior have the largest deviations, in particular the result obtained using the waveform {\it IMRPhenomD}. The cause for this deviation is currently unknown. This does serve to question the validity of these results. We also observe a disparity between the inclinations obtained between the two priors. However, the median values obtained all suggest an inclination greater than $33^{\circ}$. Its important to note that this was a single detector event which may account for the deviations in the results.

\subsection{GW200115}
This BHNS merger event was identified throughout all three detectors of the LVC network with a SNR of 11.5 \citep{abbott2021observation}. The component masses for this system are $m_1 = 5.9(+2.0,-2.5) ~\mathrm{M_{\odot}}$ and $m_2 = 1.44(+0.85,-0.28) ~\mathrm{M_{\odot}}$ \citep{abbott2021observation}. No EM counterpart to this event was detected. We present here the results using a uniform and sinusoidal prior in inclination. For the spins, as mentioned in Sec.~\ref{sec:priors} we only consider the high-spin case. The results for the analysis performed using a uniform and sinusoidal prior in inclination are given in Tables \ref{tab:GW200115_uniform} and \ref{tab:GW200115_sine}, respectively.

The results found here have some deviations with those found by the LVC. With regards to the chirp mass, the LVC finds a detector frame chirp mass of $2.58 ~\mathrm{M_{\odot}}$ \citep{abbott2021observation}. We see some deviation from this value, with the biggest deviations obtained when the waveform {\it IMRPhenomXPHM} was used. This serves to suggest that this waveform model was not that good a fit of the signal, while the results obtained using {\it IMRPhenomP} were closer to those obtained by the LVC. With regards to the inclinations obtained, we see some disagreement between the two waveform models and the two priors. The inclinations obtained by the uniform prior all settle around the median value of the interval chosen. This may be an artifact of the uniform prior. Interestingly, there is a major deviation in the values obtained by the two waveform models when a sinusoidal prior in inclination was used. This may be due to the waveform signal mismatch of {\it IMRPhenomXPHM}.

\begin{table}
    \centering
    \begin{tabular}{cccc}
    \hline
        Waveform & Inclination & Chirp mass & Mass ratio  \\
         & ($^\circ$) & ($\mathrm{M_{\odot}}$) &  \\
    \hline
       {\it IMRPhenomP} & $45.81^{+31.67}_{-35.44}$ & $2.69^{+0.00}_{-0.00}$ & $0.50^{+0.09}_{-0.12}$\\
       {\it IMRPhenomXPHM} & $ 48.15^{+28.27}_{-30.49}$ & $2.75^{+0.00}_{-0.00}$ & $0.39^{+0.12}_{-0.10}$\\
    \hline
    \end{tabular}
    \caption{Results for GW200115 using a uniform prior in inclination.}
    \label{tab:GW200115_uniform}
\end{table}

\begin{table}
    \centering
    \begin{tabular}{cccc}
    \hline
        Waveform & Inclination & Chirp mass & Mass ratio  \\
         & ($^\circ$) & ($\mathrm{M_{\odot}}$) &  \\        
    \hline
        {\it IMRPhenomP} & $60.16^{+37.24}_{-21.77}$ & $2.55^{+0.01}_{-0.01}$ & $0.27^{+0.13}_{-0.05}$\\
        {\it IMRPhenomXPHM} & $126.62^{+0.74}_{-11.45}$& $2.69^{+0.00}_{-0.00}$ & $0.50^{+0.09}_{-0.10}$\\
    \hline
    \end{tabular}
    \caption{Results for GW200115 using a sinusoidal prior in inclination.}
    \label{tab:GW200115_sine}
\end{table}

\subsection{GW190917}
This BHNS event was initially identified as a binary black hole (BBH) event by the gstLAL search pipeline \citep{abbott2023open}. The component masses involved in this merger are $m_{1} = 9.7 (+3.4, -3.9) ~\mathrm{M_{\odot}}$ and $m_2 = 2.1 (+1.1, -0.4) ~\mathrm{M_{\odot}}$, making it more likely to be a BBH event. We include it here in this study as a NSBH merger event as the secondary mass lies just below the upper bound for the theoretical largest neutron star mass $2.01 (+0.04, -0.04) \leq \mathrm{M_{TOV}}/ \mathrm{M_{\odot}} \lessapprox 2.16 (+0.17, -0.15)$ \citep{rezzolla2018using}. The measured SNR for this event was $8.3 (+0.5, -08)$ \citep{abbott2023open}. The results for GW190917 for uniform and sinusoidal priors in the inclination are given in Tables \ref{tab:GW190917_uniform} and \ref{tab:GW190917_sine}, respectively.

\begin{table}
    \centering
    \begin{tabular}{cccc}
    \hline
        Waveform & Inclination & Chirp mass & Mass ratio  \\
         & ($^\circ$) & ($\mathrm{M_{\odot}}$) &  \\       
    \hline
       {\it IMRPhenomP} & $44.63^{+30.14}_{-29.38}$ & $4.07^{+0.05}_{-0.90}$ & $0.25^{+0.01}_{-0.01}$\\
       {\it IMRPhenomXPHM} & $44.77^{+31.02}_{-30.26}$ & $4.07^{+0.05}_{-0.24}$ & $0.25^{+0.01}_{-0.01}$\\
    \hline
    \end{tabular}
    \caption{Results for GW190917 using a uniform prior in inclination.}
    \label{tab:GW190917_uniform}
\end{table}

\begin{table}
    \centering
    \begin{tabular}{cccc}
    \hline
        Waveform & Inclination & Chirp mass & Mass ratio  \\
         & ($^\circ$) & ($\mathrm{M_{\odot}}$) &  \\       
    \hline
       {\it IMRPhenomP} & $92.81^{+48.12}_{-50.42}$ & $4.06^{+0.05}_{-0.60}$ & $0.25^{+0.01}_{-0.01}$\\
       {\it IMRPhenomXPHM} & $90.52^{+53.28}_{-49.87}$& $4.07^{+0.05}_{-0.79}$ & $0.25^{+0.01}_{-0.01}$\\
    \hline
    \end{tabular}
    \caption{Results for GW190917 using a sinusoidal prior in inclination.}
    \label{tab:GW190917_sine}
\end{table}

In this instance we find that all the obtained values for the inclination tended to the median values of the intervals on the priors for inclination. This is likely due to the weak SNR but may also be an artifact of the analysis performed. Another concern is the chirp mass found in the analysis. The masses found here all tended towards $4.06 ~\mathrm{M_{\odot}}$, while LIGO found the chirp mass to be $3.7\pm 0.2 ~\mathrm{M_{\odot}}$ \citep{web:gwosc}. The lower bound of the results found in this study does include this value, however. 

\section{Discussion}
It is important to note that there is a strong degeneracy between the luminosity distance and the inclination  \citep{usman2019constraining}. As a result, the measurement of these two parameters is rather difficult through GW analysis alone. The differences in the amplitudes of the plus and cross polarizations theoretically would allow one to measure the inclination of the binary, however at small inclinations $(\iota < 45^{\circ})$ they have nearly identical amplitudes making them hard to distinguish from each other. On the other hand, at inclinations greater than $45^{\circ}$ they have much lower amplitudes. As a result, observations are biased towards face-on detection \citep{usman2019constraining}. Without an independent measurement on the luminosity distance, it becomes very difficult to constrain the inclination. This poses some problems for this study as there were no distance constraints on the progenitor binaries from observations, independent of the GW detection.

In order to determine whether it was possible to detect GRBs, if formed in the cases other than the GRB~170817A, we make the assumption that SGRBs have a viewing angle of $33^{\circ}$, which is similar or larger than the GRB jet-opening angle, based on the observations of GRB~170817A \citep{abbott2017gravitational}. Various models, however, present a range of viewing angles of SGRBs ranging from $18^{\circ}$ to $33^{\circ}$ \citep{howell2019joint}. With this constraint, we can argue that for binaries with inclinations greater than $33^{\circ}$ (or less than $147^{\circ}$), observation of the GRB would not be possible. In reality, it is more likely that in general, the viewing angle of SGRBs is smaller than $33^{\circ}$ posed in this study. This implies that being able to distinguish whether a binary is face-on or edge-on is sufficient to determine whether observation of the GRB is possible or not. The median values obtained for the inclinations of GW190425, GW200115 and GW190917, all seem to indicate that the progenitor binaries of these events may have had inclinations greater than $33^{\circ}$. This suggests that observation of any SGRB emission from these events might not have been possible. 

\begin{table}
    \centering
    \begin{tabular}{ccccc}
    \hline
        Event & $m_1$ & $m_2$ & Insprial range & Rate \\
            & ($M_\odot$)  &   ($M_\odot$)   & (Mpc) & (yr$^{-1}$) \\
    \hline
        GW170817 & $1.47^{+0.12}_{-0.10}$ & $1.26^{+0.09}_{-0.09}$ & $179^{+10}_{-10}$ & $12^{+30}_{-10}$\\
        GW190425 & $2.1^{+0.5}_{-0.4}$ & $1.3^{+0.3}_{-0.2}$ & $208^{+36}_{-28}$ &$18^{+72}_{-16}$ \\
        GW200115 & $5.9^{+2.0}_{-2.5}$ & $1.44^{+0.85}_{-0.28}$ & $304^{+102}_{-70}$ & $3^{+81}_{-3}$\\
        GW190917 & $9.7^{+3.4}_{-3.9}$ & $2.1^{+1.1}_{-0.4}$ & $410^{+119}_{-84}$ & $8^{+180}_{-8}$\\
    \hline    
    \end{tabular}
    \caption{Inspiral ranges corresponding to the component masses of the GW merger events in this study. We also present the event rates corresponding to the calculated inspiral ranges, using the local rates of BNS and BHNS from \citet{burns2020neutron}.}
    \label{tab:rate}
\end{table}

Next, we calculate the rate of occurrence of GW events like the ones that we have analysed. The range for a BNS or BHNS system with component masses $m_1$ and $m_2$ can easily be found using \textsc{GWINC},\footnote{\url{ https://pypi.org/project/gwinc/}} a python package for processing noise budgets for ground based interferometers \citep{barsotti2018updated}. Using the \textsc{GWINC} parameters found in  for the noise budget of the LIGO network, we find the inspiral range that are listed in Table~\ref{tab:rate} against the GW events. The corresponding rates are calculated using local rates given in \citet{burns2020neutron}. Note that the detection rate of the two BNS events, GW170817 and GW190425, is $\ge 2$~yr$^{-1}$. The lowest rate of the two BHNS events, on the other hand, is consistent with zero.   


\begin{figure}
    \centering
    \includegraphics[scale=0.25]{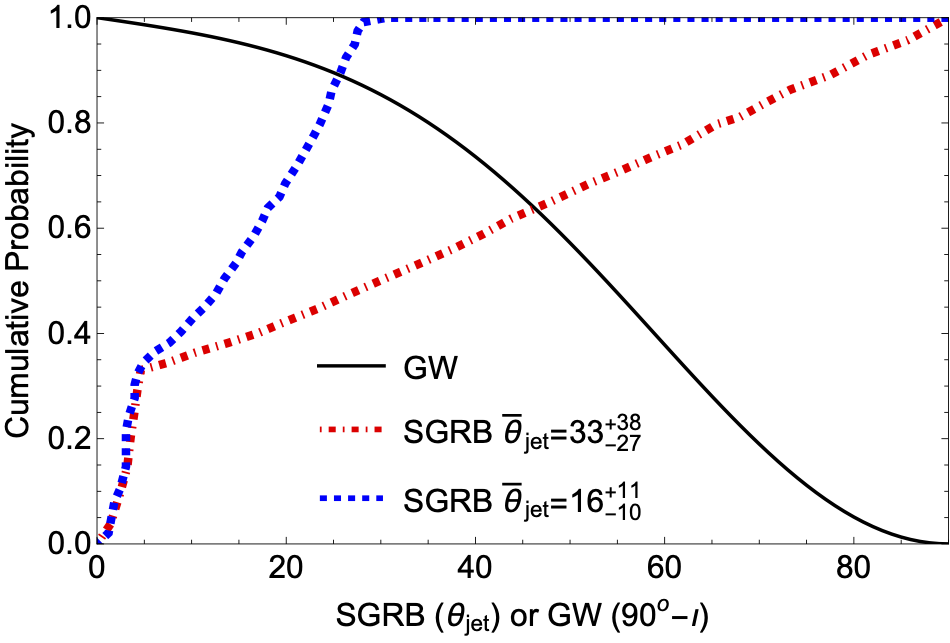}
    \caption{Cumulative probability distributions (CPDs) for SGRB emission and GWs from BNS mergers, as functions of the SGRB jet opening angle ($\theta_{\rm jet}$) and $90^\circ -\iota$; where $\iota$ is the binary inclination angle, respectively. The CPDs for SGRB have been adapted from \citet{Fong2015ApJ...815..102F}, where the maximum jet-opening angle used was $30^\circ$ (blue dashed curve) and $90^\circ$ (red dotted-dash curve). The CPD for GW has been calculated from equation~(\ref{eq:GW_pdf}).}
    \label{fig:GRB_GW}
\end{figure}

To then find what the chances of a GW/GRB joint detection from a BHNS merger, is not so easy. The fraction of successful GRBs emitted from BHNS mergers is currently unknown. As a result, estimates for joint detection mainly focus on BNS mergers. It is quite likely that the lack of an EM counterpart to the two BHNS mergers analysed in this study, is due to the fact that the GRB was not successfully produced in the merger. However, assuming that BHNS mergers can produce successful GRBs, we could potentially find an upper bound on what could be the GRB/GW joint detection rate from BHNS mergers using what we know from BNS mergers. GW emission is omnidirectional but not isotropic. The observed inclination angle probability distribution of GWs is given by \citep{schutz2011networks}
\begin{equation}
    \rho_{\rm GW-detected}(\iota) = 0.002656(1 + 6\cos^2(\iota) + \cos^4(\iota))^{3/2}\sin(\iota)
    \label{eq:GW_pdf}
\end{equation}
where $\iota$ is the orbital inclination angle. Figure~\ref{fig:GRB_GW} shows the cumulative probability distribution (CPD) from equation~(\ref{eq:GW_pdf}) as a function of $90^\circ - \iota$, where $\iota = 90^\circ$ is face-on and $\iota = 0^\circ$ is edge-on orientation of the orbital plane from an observer's perspective. Figure~\ref{fig:GRB_GW} also shows CPDs of SGRBs as functions of the jet-opening angle $\theta_{\rm jet}$ for two values of the average jet opening angle ${\bar \theta}_{\rm jet} = 16^\circ$($+11^\circ$, $-10^\circ$) and $33^\circ$($+38^\circ$, $-27^\circ$), as derived by \citet{Fong2015ApJ...815..102F} from observed data. Therefore, the probability of GW/GRB joint detection is roughly 1/9 BNS or BHNS mergers for ${\bar \theta}_{\rm jet} = 16^\circ$($+11^\circ$, $-10^\circ$), which is similar to the number found by \citet{burns2020neutron}. For ${\bar \theta}_{\rm jet} = 33^\circ$($+38^\circ$, $-27^\circ$), the joint detection probability is roughly 1/2 BNS or BHNS mergers. Non-detection of a GRB in the BNS event GW190425 or the BHNS events GW200115 and GW190917 is consistent with both values of the average SGRB jet opening angle ${\bar \theta}_{\rm jet}$. Non-detection of an SGRB from a number of detected BNS or BHNS mergers in future can therefore put constraints on the SGRB jet opening angle, assuming an SGRB is formed in all those mergers.     


%


\section{Conclusions}
The conclusions presented in this study are rather tentative, due to concerns over large uncertainties in our results stemming from the fact that all GW event concerned in this study except for GW170817 were weak detections. There are relatively large uncertainties in determining the orbital inclination angle that overlaps with the $33^{\circ}$ GRB jet opening angle. This is most notable in Table~\ref{tab:GW190425_LS}. This displays an interesting disparity between the low-spin and high-spin priors used for GW190425. It's important to note that this was a single detector event which impacts the overall accuracy of the analysis that can be performed. We see similar disparities between priors when we look at the results obtained for events GW200115 and GW190917. The results obtained using a uniform prior all converge around the median value of the interval chosen, while the inclinations obtained through a sinusoidal distribution in inclination tends to give different values. Particularly in the case of GW200115, the inclination angles are quite different between the waveform models used. The reason for this disparity is again weak detection of the event. Taking into account these concerns, the results obtained in this study can only suggest that these binaries were orientated such that observation of the emitted GRB was not possible. In order to obtain a more accurate measurement one needs an independent means of placing a constraint on the luminosity distance of the GW sources. While tentative, we believe our results are interesting enough for further investigation with more GW and GRB data in future.  

\section*{Acknowledgements}

We thank the Referee, Joshua Wood, for carefully reading the manuscript and providing feedback that has improved the presentation. This work was made possible with grants from the National Research Foundation (NRF), South Africa,  BRICS STI programme; the National Institute for Theoretical Computational Sciences (NITheCS), South Africa; and the South African Gamma-ray Astronomy Programme (SA-GAMMA). LM was supported by a GES MSc bursary from the University of Johannesburg (UJ) and an NRF MSc bursary. This work used High Performance computing clusters at UJ and at the University of Witwatersrand. Without access to these computational resources, this study would not have been feasible. 


\section*{Data Availability}
All data used in this study is publicly available at \cite{web:gwosc}. The cleaned data used to analyse GW170817 is available at \cite{web:cleandata}. Information concerning the Bayesian libraries used in this study can be found at \cite{ashton2019bilby}.



\bibliographystyle{mnras}
\bibliography{example} 








\bsp	
\label{lastpage}
\end{document}